\documentclass[twoside,11pt]{article}

\usepackage[a4paper,lmargin=2.5cm,rmargin=2cm,tmargin=1cm,bmargin=1cm,includehead,includefoot]{geometry}
\usepackage[german,english]{babel}
\usepackage{natbib}
\usepackage[ruled,lined,boxed,linesnumbered]{algorithm2e}
\usepackage{url}
\usepackage{subfig}
\usepackage{fancyhdr}
\usepackage{array}
\usepackage{amsmath} % needed for subequations
\usepackage{amssymb}
\usepackage{mathptmx}
\usepackage{multirow}
\usepackage[scaled=.90]{helvet}
\usepackage{times}
\usepackage{graphicx}
\usepackage{hyperref}

\hypersetup{colorlinks=true, breaklinks=true, pagebackref=true,
  urlcolor=blue, linkcolor=blue,anchorcolor=blue,citecolor=blue,
  pdfpagemode = UseNone, %FullScreen, %UseThumbs, %UseOutlines,
  pdfauthor = {},
  pdftitle = {},
  pdfsubject = {},
  pdfkeywords = {}
}

\DeclareGraphicsExtensions{.jpg,.pdf,.mps,.png}
\graphicspath{{img/} {./}} %put all figures in these dirs

\fancypagestyle{plain}{%
\fancyhf{}
\lfoot{}
\rfoot{\thepage}
\chead{}
\rhead{}
\lhead{}

}
\title{Efficient Computation of Mean Truncated Hitting Times \\ on Very Large Graphs}
\author{Joel Lang and James Henderson\\
  Computational Learning and Computational Linguistics Research Group\\
  University of Geneva\\
}
\date{5.12.2012}

\begin{document}

\maketitle
\thispagestyle{empty}

% Abstract
\begin{abstract}
Previous work has shown the effectiveness of random walk hitting times as a measure of dissimilarity in a variety of graph-based learning problems such as collaborative filtering, query suggestion or finding paraphrases. However, application of hitting times has been limited to small datasets because of computational restrictions. This paper develops a new approximation algorithm with which hitting times can be computed on very large, disk-resident graphs, making their application possible to problems which were previously out of reach. This will potentially benefit a range of large-scale problems.
\end{abstract}
% keywords

\section{Introduction}
Efficient algorithms for graph-based learning have become central towards solving a range of large-scale learning problems. Prominent examples are ranking of web pages \citep{authoritative_sources_in_a_hyperlinked_environment,the_pagerank_citation_ranking_bringing_order_to_the_web}, collaborative filtering \citep[e.g.,][]{a_random_walks_perspective_on_maximizing_satisfaction_and_profit} and general-purpose methods for semi-supervised classification \citep[e.g.,][]{semi_supervised_learning_using_gaussian_fields_and_harmonic_functions}. Typically, graph vertices represent instances (e.g.\ webpages) and an edge indicates that two vertices are in some sense \emph{close} or \emph{similar}. Many of these problems involve graphs with hundreds of millions or billions of vertices and therefore call for methods which scale to such data quantities and in particular for methods which can be parallelized and executed on a cluster of machines.

Apart from the issue of scalability, a fundamental question underlying graph-based methods is how the information encoded in the edge weights of a graph can be exploited most effectively in order to transfer information between graph vertices, for example to propagate labels for classification tasks and similarly to propagate rank for ranking tasks. Interestingly, many successful methods can be related to the notion of graph random walks. In fact, \emph{all} of the examples cited above either directly or indirectly correspond to computing a random walk on a graph. Moreover, previous work provides evidence that (dis-)similarity measures\footnote{Here simply meaning a function which measures \\(dis-)similarity between vertices on some scale.} based on random walk \emph{hitting times} are particularly effective for many problems. For example, \cite{query_suggestion_using_hitting_time} show that hitting times outperform alternative methods such as topic-sensitive pagerank \citep{topic_sensitive_pagerank_a_context_sensitive_ranking_algorithm_for_web_search} for query suggestion, \cite{random_walk_computation_of_similarities_between_nodes_of_a_graph_with_application_to_collaborative_recommendation} sucessfully employ hitting times for collaborative filtering, \cite{hitting_the_right_paraphrase_in_good_time} generate paraphrases using hitting time based scoring, \cite{clustering_using_a_random_walk_based_distance_measure} use a hitting time based measure for clustering and \cite{shape_representation_and_classification_using_the_poisson_equation} use hitting times for computing shape representations.

The mean hitting time of some vertex is the expected number of steps it takes for a random walk to reach that vertex, starting from some start vertex. One property which makes hitting times an appropriate dissimilarity measure in the aforementioned applications is their ability to reflect the overall connectivity structure of the graph, in contrast to measures such as the shortest path between two vertices. The hitting time will decrease when the number of paths from the start vertex to the target vertex increases, when the length of paths decreases or when the likelihood (weights) of paths increases. These properties are particularly important for problems where the graph edges encode a transitive (but not necessarily symmetric) relationship and must be assumed to be merely a sample of all plausible edges, possibly perturbed by noise. Consider for example the web graph, where a link from page $a$ to page $b$ will often indicate that $b$ is relevant to $a$. This relationship is transitive  (to some degree) and a page will link only to a sample of relevant pages. In such a setting, the hitting time for some vertex/page $c$ and start vertex $a$ will capture how relevant page $c$ is for $a$ and reflect all the evidence encoded in the graph about the relationship between the two.

So far, one drawback of methods based on hitting times compared to other methods based on random walks has been that computing the expected hitting times for a fixed start vertex to all other vertices is expensive, especially when using the standard iterative algorithm, as this requires computing a dense $n \times n$ matrix, where $n$ is the number of vertices (details will follow below). For large graphs such as those mentioned above, computing this matrix is infeasible. Therefore, previous work has either investigated hitting times only on relatively small-scale problems with at most thousands of vertices \citep[e.g.,][]{random_walk_computation_of_similarities_between_nodes_of_a_graph_with_application_to_collaborative_recommendation} or has been concerned with finding approximation algorithms which find upper and lower bounds and make use of sampling \citep{fast_incremental_proximity_search_in_large_graphs,a_tractable_approach_to_finding_closest_truncated_commute_time_neighbors_in_large_graphs}. However, none of the proposed methods will scale to very large graphs which reside on disk rather than main memory.

This paper resolves the scalability problem by developing an approximation algorithm for computing mean truncated hitting times. The algorithm is space and runtime efficient, storing only $3$ floating point numbers per vertex (in addition to the graph) and its runtime is linear in the number of edges. With our algorithm it becomes possible to compute hitting times for large-scale problems, where they could previously not be applied. Section \ref{sec:background_and_previous_work} will provide some background and describe previous work. Then in Section \ref{sec:efficient_computation_of_hitting_times} we describe our approximation algorithm. In Section \ref{sec:experiments} we will provide an empirical assessment of approximation accuracy.

\section{Background and Previous Work}
\label{sec:background_and_previous_work}

A random walk on a graph with vertices $V=\{1 \hdots n \}$ is a discrete-time Markov chain $(X_t)_{t \in \mathbb{Z}^+}$, defined in terms of an initial distribution $\lambda$ over $V$ and a (row-)stochastic matrix $P$ which captures the transition probabilities between vertices: $P_{ij}=p(X_{t+1}=j|X_t=i)$. For many applications, the transition matrix is a sparse matrix derived from the edge weights of the graph, which can be either directed or undirected.

The hitting time for some vertex $j \in V$ is the random variable $\Delta_{\lambda,j}$ over $\mathbb{Z}^+ \cup \{ \infty \}$ which measures how long it takes until the random walk first hits $j$:
\begin{equation*}
    \Delta_{\lambda,j} = \inf \{t \geq 0 | X_t=j\}
\end{equation*}
where, as in \cite{markov_chains}, the infimum of the empty set is $\infty$. To simplify the discussion, we will initially assume that $\lambda$ is a distribution with unit mass on vertex $i$, i.e., $\lambda=\delta_i$, and we will simply write $\Delta_{\delta_i,j}=\Delta_{ij}$. The vertex $i$ will be called the \emph{start vertex}.

The $T$-truncated hitting time, introduced by \cite{a_tractable_approach_to_finding_closest_truncated_commute_time_neighbors_in_large_graphs}, is defined as
\begin{equation*}
    \Delta^{(T)}_{ij} = min(\Delta_{ij},T)
\end{equation*}
Clearly, as $T$ increases the truncated hitting time approaches the untruncated hitting time and asymptotically they are the same. For practical purposes, when hitting times are used for defining a (dis-)similarity measure it is usually sufficient or even superior \citep{a_tractable_approach_to_finding_closest_truncated_commute_time_neighbors_in_large_graphs} to compute truncated hitting times instead of untruncated hitting times, where typical values for $T$ are between $10$ and $20$. The choice can be based on the mixing rate of the Markov chain, i.e., the rate at which the distribution over vertices converges towards the stationary distribution.

We are interested in computing the expected value of $\Delta^{(T)}_{ij}$, i.e., the mean truncated hitting time:
\begin{equation}
    h^{(T)}_{ij} = E[\Delta^{(T)}_{ij}] = \sum_{t=0}^{T} t P[\Delta^{(T)}_{ij}=t]
    \label{equ:direct_definition}
\end{equation}
The approach taken in \cite{a_tractable_approach_to_finding_closest_truncated_commute_time_neighbors_in_large_graphs} is based on the following recursive definition of hitting times, which is equivalent with the definition above:
\begin{equation}
    h^{(T)}_{ij} = \begin{cases}
	0 \hspace{3 em} i=j \vee T=0
	\\
	1 + \sum_{k \in V} p_{ik} h^{(T-1)}_{kj}
\end{cases}
\label{equ:recursive_definition}
\end{equation}
The problem with this approach is that, while computing $h^{(T)}_{ij}$ for all $i$ and fixed $j$ is relatively straightforward, computing $h^{(T)}_{ij}$ for all $j$ and fixed start vertex $i$ is computationally expensive and requires computing the full $n \times n$ matrix of mean truncated hitting times at intermediate steps of the computation. This can be seen from Equation \ref{equ:recursive_definition}, where in the second case $h^{(T-1)}_{kj}$ is required for all neighbors $k$ in order to compute $h^{(T)}_{ij}$, and in order to compute $h^{(T-1)}_{kj}$ we need the $T-2$ truncated hitting times for all of $k$'s neigbors, and so on. For large $n$ computing a dense $n \times n$ matrix is intractable, which is why \cite{a_tractable_approach_to_finding_closest_truncated_commute_time_neighbors_in_large_graphs} have proposed a pruning scheme with which hitting times can be computed approximately.

Specifically, they derive upper and lower bounds which are precomputed and stored for each pair of close neighbors. The set of close neighbors is determined by iterative expansion (see \cite{a_tractable_approach_to_finding_closest_truncated_commute_time_neighbors_in_large_graphs} for details). Once the bounds for all close neighbors have been precomputed they can then be queried in order to compute bounds on the hitting time for an arbitrary pair of vertices. We will briefly review here how these bounds are computed.

Let $\mathcal{N}(i)$ denote the set of direct neighbors of $i$ (reachable within one step), let $\mathcal{C}(j)$ denote a given set of vertices with short paths leading to vertex $j$ and let $\mathcal{B}(j) \subset \mathcal{C}(j)$ denote a set of boundary vertices, which also have paths leading to vertices outside $\mathcal{C}(j)$. For a start vertex inside the close neighbors $i \in \mathcal{C}(j)$ the upper bound $\overline{h}^{(T)}_{ij}$ is computed as
\begin{align*}
    \overline{h}^{(T)}_{ij} & = 1 + \sum_{k \in \mathcal{C}(j) \cap \mathcal{N}(i)} P_{ik} \overline{h}^{(T-1)}_{kj}
    \\ & + (1 - \sum_{k \in \mathcal{C}(j) \cap \mathcal{N}(i)} P_{ik}) (T-1)
\end{align*}
and the lower bound $\underline{h}^{(T)}_{ij}$ is
\begin{align*}
    \underline{h}^{(T)}_{ij} & = 1 + \sum_{k \in \mathcal{C}(j) \cap \mathcal{N}(i)} P_{ik} \underline{h}^{(T-1)}_{kj}
    \\ & + (1 - \sum_{k \in \mathcal{C}(j) \cap \mathcal{N}(i)} P_{ik}) (1 + \min_{l \in \mathcal{B}(j)} \underline{h}^{(T-2)}_{lj})
\end{align*}
Once these bounds have been computed for each pair of close neighbors, the upper bound for a start vertex outside the close neighbors $i \notin \mathcal{C}(j)$ is simply $T$ and the lower bound is
%given by
\begin{align*}
    \underline{h}^{(T)}_{ij} & = 1 + \min_{k \in \mathcal{C}(j)} \underline{h}^{(T-1)}_{kj}
\end{align*}

While their algorithm can help reduce the storage requirements, storing all pairs of close neighbors and precomputing bounds for them is likely to be intractable for very large graphs and in the worst case requires $O(n^2)$ space and time. To improve performance \cite{fast_incremental_proximity_search_in_large_graphs} resort to sampling for computing the hitting times from a start vertex, which is combined with the pruning scheme into an algorithm for computing the approximate top-$k$ nearest commute time neighbors for a given query vertex. However, a sampling-based approach becomes inefficient when the graph does not fit into main memory, because repeated random access to disk will result in thrashing.

In the following section we will present an algorithm for approximately computing $h^{(T)}_{ij}$ for all $j$ and fixed start vertex $i$ which runs in $O(T|E|+Tn)$ time, where $|E|$ is the number of graph edges, and which stores only $3n$ additional floating point numbers as opposed to the $n^2$ floating point numbers stored in the conventional approach. Moreover, the algorithm is straightforward to parallelize for example within the map-reduce paradigm. Therefore, our approach is well-suited for disk-resident graphs, in contrast to sampling-based approaches.

\section{Efficient Computation of Hitting Times}
\label{sec:efficient_computation_of_hitting_times}

Our approach is based on the direct definition of hitting times given by Equation \ref{equ:direct_definition}, rather than the recursive definition in Equation \ref{equ:recursive_definition}. The t-step transition matrix is written as $P^t$ and accordingly the t-step transition probabilities are written as $P^t_{ij}$. Assuming that the start vertex $i$ is fixed, we must compute $P[\Delta^{(T)}_{ij}=t]$ for each $j$ and $t$, in order to evaluate the sum in Equation \ref{equ:direct_definition}. For compactness we will write
\begin{equation*}
    P^{*t}_{ij} = P[\Delta^{(T)}_{ij}=t] \hspace{2 em} (t<T).
\end{equation*}
For $T$ we have
\begin{equation*}
    P^{*T}_{ij} = 1 - \sum_{t=0}^{T-1} P^{*t}_{ij}
\end{equation*}

For the non-trivial case where $i\neq j$ we can write
\begin{equation}
    P^{*t}_{ij} = P^t_{ij} - \sum_{k=1}^{t-1} P^{*k}_{ij} P^{t-k}_{jj}
    \label{equ:starting_point}
\end{equation}

Proof: Let $S$ be the set of all paths of length $t$ starting at $i$. Let $S_j^k \subset S$ be the set of paths which pass $j$ the first time after $k$ steps and end in $j$. The set $S_j \subset S$ of \emph{all} paths ending in $j$ is then given by $S_j = \cup_{k=1}^{t} S_j^k$. Since the sets $S_j^k$ are mutually disjoint we have
\begin{align*}
    P[S_j] & = P[\cup_{k=1}^{t} S_j^k] = \sum_{k=1}^{t} P[S_j^k] \\
    P[S_j^t] & = P[S_j] - \sum_{k=1}^{t-1} P[S_j^k]
\end{align*}
Now substituting $P[S_j]=P^t_{ij}$, $P[S_j^t]=P^{*t}_{ij}$ and $P[S_j^k]=P^{*k}_{ij} P^{t-k}_{jj}$ for $k<t$ we obtain Equation \ref{equ:starting_point}.

Consider again Equation \ref{equ:starting_point}: computing the values $P^t_{ij}$ for all $j$ and fixed $i$ can be done in a space-efficient manner, by iteratively multiplying the transition matrix with the state-distribution vector at $t-1$. In contrast, computing the values $P^t_{jj}$ requires us to compute all of the diagonal entries of $P^t$, which if done exactly requires computing the full t-step transition matrix for each $t<T$. Again, for large $n$ this intractable.

We could at this point resort to sampling in order to estimate the $P^t_{jj}$ with any desired accuracy with high probability. Let $X^t_i$ be the Bernoulli random variable which indicates whether a random walk starting at $i$ hits $i$ after $t$ steps, with probability $P[X^t_i=1]=E[X^t_i]=p_{ii}$. We can obtain independent samples $\{x^t_{i0}\hdots x^t_{iL}\}$ of this random variable by sampling $L$ (independent) random walks, each of length $T$ since we are interested in all $t<T$. We will write $\mu^t_i = \frac{1}{L}\sum_{l=1}^L x^t_{il}$ for the empirical mean obtained from the sample. Then, using Hoeffding bounds we have
\begin{equation*}
        P[|\mu^t_i-p^t_{ii}| \geq \epsilon] \leq 2 \exp \left(-2 \epsilon^2 L\right)
\end{equation*}
Thus in order to have an $\epsilon$-correct estimate of $p^t_{ii}$ with at least probability $1-\rho$ we must sample at least $L=\frac{1}{2\epsilon^2} \log \left( \frac{2}{\rho} \right)$ random walks for each vertex. Note that approximation errors to $p^t_{ii}$ will be reduced when computing $P^{*t}_{ij}$ according to Equation \ref{equ:starting_point} since they are multiplied by $P^{*k}_{ij}\leq 1$ and errors will tend to be cancelled out by the sum. On the other hand, they may be amplified by factor $t$ when computing the actual expectation according to Equation \ref{equ:direct_definition}.

%An advantage of this sampling method over \cite{fast_incremental_proximity_search_in_large_graphs} is that the $p^t_{ii}$ can be precomputed once for all vertices and then reused when computing the hitting times for \emph{any} start vertex $i$, assuming that storing the $Tn$ values is tractable. Thus if we want to compute exact hitting times for multiple start vertices our method will be bring a considerable speedup.

As was mentioned above, a major disadvantage of a sampling-based approach is that it is inefficient when the graph does not fit into main memory, because repeated random access to disk will result in thrashing. In the following we will therefore avoid sampling and discuss an alternative algorithm with which approximate hitting times can be computed very efficiently. In practice there are many situations where obtaining a loose estimate of hitting times is sufficient.  Our algorithm essentially renders sampling the $P^t_{jj}$ unnecessary and therefore avoids the most expensive part of the computation of the sampling approach.

\subsection{The Approximation}
\label{sec:approximation}

\begin{algorithm}[t]
\SetKwInOut{Input}{input}
\SetKwInOut{Output}{output}
\caption{Iterative, space-efficient computation of approximate mean truncated hitting times}
\Input{vertices $V$ \\ transition matrix $P$ (sparse) \\ start vertex $i$}
\Output{$h_j^{(T)} \approx h^{(T)}_{ij}$ for each $j \in V$}
\BlankLine
$h^{(0)} \leftarrow \mathbf{0}$
\\
$p^{(0)} \leftarrow \delta_i$
\\
$f^{(0)} \leftarrow \mathbf{1} - p^{(0)}$
\\
\For{$t = 1 \hdots T-1$}{
        $p^{(t)} \leftarrow P^\top p^{(t-1)}$
        \\
        $h^{(t)} \leftarrow h^{(t-1)} + t (p^{(t)} \circ f^{(t-1)})$
        \\
        $f^{(t)} \leftarrow  f^{(t-1)} \circ (\mathbf{1}-p^{(t)})$
}

$h^{(T)} \leftarrow h^{(T-1)} + T f^{(T-1)}$
\label{alg:iterative_algorithm}
\end{algorithm}

The approximation is based on the following equation
\begin{equation}
    P^{*t}_{ij} = P[(X_t=j) \wedge (X_{t-1}\neq j) \hdots \wedge (X_{1}\neq j)| X_0=i]
    \label{equ:starting_equation}
\end{equation}
We can then make the simplifying assumption that the events occurring in the conjunction are (almost) independent. The larger the graph, the larger $t$ and the larger the mixing rate of the Markov chain, the more accurately this assumption holds. Intuitively, for many graphs, especially those representing `real-world' application data, knowing that we are not in state $j$ at some point of the Markov chain will provide only very little information on average about where we will be in the following states. This leads to the approximation:
\begin{equation*}
    P^{*t}_{ij} \approx P^{t}_{ij} \prod_{k=0}^{t-1} (1-P^{k}_{ij})
\end{equation*}
Likewise, for $t=T$ we can use
\begin{equation*}
    P^{*T}_{ij} \approx \prod_{k=0}^{T-1} (1-P^{k}_{ij})
\end{equation*}

Putting everything together results in the following approximation scheme
\begin{align*}
    h^{(T)}_{ij} ~\approx~ & \sum_{t=0}^{T-1} t \left[ P^t_{ij} \prod_{k=0}^{t-1} (1-P^k_{ij}) \right]
    %\\ &
    ~+~ T \left[ \prod_{k=0}^{T-1} (1-P^k_{ij}) \right]
    \label{equ:computed_definition}
\end{align*}

%If we have upper bounds $\overline{P}^t_{jj}$ and lower bounds $\underline{P}^t_{jj}$ we can use them to compute bounds upper and lower bounds for $P^{*t}_{ij}$:
%\begin{align*}
%    P^{*t}_{ij} & \leq  \overline{P}^{*t}_{ij} = \min(P^t_{ij} - \sum_{k=1}^{t-1} \underline{P}^{*k}_{ij} \underline{P}^{t-k}_{jj},1)
%    \\
%    P^{*t}_{ij} & \geq  \underline{P}^{*t}_{ij} = \max(P^t_{ij} - \sum_{k=1}^{t-1} \overline{P}^{*k}_{ij} \overline{P}^{t-k}_{jj},0)
%\end{align*}

Based on this equation we can then compute $h^{(T)}_{ij}$ for each $j$ and fixed $i$ according to Algorithm~1. The algorithm proceeds by iteratively computing the terms occurring in the equation above. It maintains three vectors $h$, $p$ and $f$. The vector $h^{(t)}$ stores the sum up to term $t$ for each possible target vertex $j$, s.t.\ the final result $h^{(T)}$ corresponds to the vector of approximate mean $T$-truncated hitting times of a random walk starting at $i$. The vector $p^{(t)}$ stores the distribution over vertices of the random walk after $t$ steps, i.e., it stores the $i-th$ row of the t-step transition matrix $P^t$. Finally, the vector $f^{(t)}$ stores the product $\prod_{k=0}^{t} (1-P^k_{ij})$ for each $j$. We have used the notation $a \circ b$ to denote the component-wise multiplication of two (same-length) vectors $a$ and $b$.

%The main advantage of this algorithm is that we do not need to store a dense $n \times n$ matrix as we would using the recursive definition in Equation \ref{equ:recursive_definition}, which requires computing the full matrix $H^{(t)}$ of hitting times for intermediate values of $t$. In particular, we do not need to store the full $t$-step transition matrix $P^t$ but only a single row, i.e., the current distribution over vertices.

The space required by our algorithm is simply the space for the three vectors $h$, $p$ and $f$, giving a total of $3n$ floating point numbers (excluding storage of the graph, which resides on disk). The main computational load of the algorithm stems from the matrix-vector multiplication in Line 5 between the transition matrix $P$ and the current distribution $p$. If $P$ is sparse, we can resort to sparse matrix-vector multiplication methods, which requires computing as many multiplications as there are non-zero components in $P$. Typically, $P$ has a non-zero component for each edge of the graph $G=(V,E)$, so each iteration requires $|E|+3n$ multiplications and $|E|+2n$ additions. The total runtime is therefore $O(T|E|+Tn)$. Importantly, devising a map-reduce version of the algorithm is straightforward, since  matrix-vector multiplication can be implemented as a map-reduce operation \citep[see][]{mining_of_massive_datasets}.
Finally, note that the algorithm can be easily applied in the case where, instead of a single start vertex, one is given a distribution $\lambda$ over start vertices. All that needs to be changed is the initialization $p^{(0)} \leftarrow \lambda$.

\subsection{Higher-Order Approximations}

Our approximation was based on Equation \ref{equ:starting_equation}, where we assumed (approximate) independence between events in the conjunction. While this results in a particularly efficient algorithm, we can improve the approximation accuracy by weakening this assumption and instead assuming $d$-th order Markovian dependencies between the events. For example for order $d{=}1$ we have
\begin{align*}
    P^{*t}_{ij} = & P[(X_t=j) \wedge (X_{t-1}\neq j) \hdots \wedge (X_{1}\neq j)| X_0=i] \\
    \approx &  P[X_t=j|X_{t-1}\neq j,X_0=i]  \hdots P[X_{1}\neq j| X_0=i] \\
    = & (P^{t}_{ij}-P^{t}_{ijj}) \prod_{k=2}^{t-1} \left(1-\frac{P^{k-1}_{ij}-P^{k}_{ijj}}{1-P^{k}_{ij}}\right)
\end{align*}
where $P^{t}_{ijj}=P(X_t=j,X_{t-1}=j|X_0=i)$. Thus for order $d{=}1$ this improves the approximation accuracy without changing the asymptotic space and runtime complexity of our algorithm.  However, for $d>1$ it would in general require computing dense $n \times n$ matrices.

\section{Approximation Accuracy}
\label{sec:experiments}

\begin{table*}[!t]
\begin{center}
%\subfloat[Scores for the baseline which assigns arguments to clusters based on their syntactic function.]{
\begin{tabular}{|l|l|l|l|l|l|l|l|l|l|} \hline
& \multicolumn{3}{c|}{Sparse 1} & \multicolumn{3}{c|}{Sparse 2}& \multicolumn{3}{c|}{Dense}\\
    	& 10 & 100 & 1000 & 10 & 100 & 1000 & 10 & 100 & 1000 \\
    \hline
    \hline
	avg err & $0.0433$ & $0.0003$ & $0.0001$ & $0.0423$ & $0.0004$ & $0.0001$ & $0.0134$ & $0.0002$ & $0.0000$
	\\
	max err & $ 0.2863$ & $0.0122$ & $0.0269$ & $0.2621$ & $0.0140$ & $0.0227$& $0.0420$ & $0.0005$ & $0.0000$
	\\	
    \hline
	avg inv & $0.0153$ & $0.0049$ & $0.0036$  & $0.0163$ & $0.0021$ & $0.0016$ & $0.0512$ & $0.0110$ & $0.0013$
	\\
	max inv & $0.0422$ & $0.0080$ & $0.0041$ & $0.0430$ & $0.0041$ & $0.0019$& $0.1156$ & $0.0159$ & $0.0015$
	\\	
  \hline
  \end{tabular}
\end{center}
\caption{Approximation accuracy on small synthetic graphs. \label{tab:approximation_accuracy}}
\end{table*}

In this section we will demonstrate that our approximation algorithm empirically results in accurate estimates of hitting times and will usually induce a ranking very close to the one produced by exact hitting times. Here we will conduct experiments on small synthetic graphs on which hitting times are exactly computable.

Specifically we will use both sparse and dense directed graphs with $10$, $100$ and $1000$ vertices respectively. For the sparse graphs we will generate $20$, $1000$ and $10000$ edges respectively. Given the number of vertices and edges, the first type of sparse graph (SP1) is generated by first randomly sampling an incoming and outgoing edge for each vertex and then sampling additional edges uniformly at random until the number of total edges is reached (if a sampled edge is already present, the sampling step is repeated). The second type of sparse graph (SP2) is generated by again first sampling an incoming and outgoing edge for each vertex. Then further edges are added by first sampling the target vertex of an edge, whereby the probability of choosing a vertex is proportional to the number of incoming edges it already possesses and then sampling a source vertex uniformly at random. This tends to accumulate edges at certain vertices and results in a different type of sparse graph. Again, the process is repeated until the desired number of edges is reached. In both cases edges receive unit weights. For generating the dense graph (DEN) we create a fully connected, directed graph and sample weights from a uniform distribution.

We compute the following scores: the average and maximum relative error of the approximation computed over all vertex pairs. If $h_{ij}$ is the exact value and $\hat{h}_{ij}$ is the approximation then the relative error is defined as $|\frac{h_{ij}-\hat{h}_{ij}}{h_{ij}}|$. We will also consider the rankings generated by the exact and approximate hitting times and compute the proportion of vertex pairs which are ranked differently in the two rankings, i.e., the relative number of inversions in the approximate ranking compared to the exact ranking. Small values indicate that the rankings are similar and a random ranking would result in around $50\%$ inversions. We consider the exact and approximate rankings for each possible start vertex and again report the average and maximum inversions. For each graph type we generate $30$ random graphs and report the aggregate scores (average and maximum) over all of these graphs in Table \ref{tab:approximation_accuracy}.

\subsubsection{Results}

Results are shown in Table \ref{tab:approximation_accuracy}. As expected, the approximation accuracy is higher for dense graphs than for sparse graphs, because the mixing rate of the Markov chain is higher and therefore the independence assumption underlying our approximation is less violated. The results also confirm that the relative error decreases with the number of vertices and is already quite small even for small graphs. As in practice our algorithm will be applied to graphs several orders of magnitudes larger, we can expect very high approximation accuracies on these graphs. Moreover, the ranking induced by the approximate hitting times are nearly identical to those produced by exact hitting times, and thus in many applications will constitute a valid replacement.

%----------------------------------------------------------%
\bibliographystyle{plainnat}
\bibliography{c:/research/Master}

\begin{thebibliography}{14}
\providecommand{\natexlab}[1]{#1}
\providecommand{\url}[1]{\texttt{#1}}
\expandafter\ifx\csname urlstyle\endcsname\relax
  \providecommand{\doi}[1]{doi: #1}\else
  \providecommand{\doi}{doi: \begingroup \urlstyle{rm}\Url}\fi

\bibitem[Brand(2005)]{a_random_walks_perspective_on_maximizing_satisfaction_and_profit}
M.~Brand.
\newblock {A Random Walks Perspective on Maximizing Satisfaction and Profit}.
\newblock In \emph{{Proceedings of the SIAM International Conference on Data
  Mining}}, 2005.

\bibitem[Fouss et~al.(2007)Fouss, Pirotte, Renders, and
  Saerens]{random_walk_computation_of_similarities_between_nodes_of_a_graph_with_application_to_collaborative_recommendation}
F.~Fouss, A.~Pirotte, J.~Renders, and M.~Saerens.
\newblock {Random-Walk Computation of Similarities between Nodes of a Graph
  with Application to Collaborative Recommendation}.
\newblock \emph{IEEE Transactions on Knowledge and Data Engineering},
  19\penalty0 (3):\penalty0 355--369, 2007.

\bibitem[Gorelick et~al.(2006)Gorelick, Galun, Sharon, Basri, and
  Brandt]{shape_representation_and_classification_using_the_poisson_equation}
L.~Gorelick, M.~Galun, E.~Sharon, R.~Basri, and A.~Brandt.
\newblock {Shape Representation and Classification Using the Poisson Equation}.
\newblock \emph{{IEEE Transactions on Pattern Analysis and Machine
  Intelligence}}, 28\penalty0 (12):\penalty0 1991 --2005, 2006.

\bibitem[Haveliwala(2003)]{topic_sensitive_pagerank_a_context_sensitive_ranking_algorithm_for_web_search}
T.~Haveliwala.
\newblock {Topic-sensitive PageRank: a Context-sensitive Ranking Algorithm for
  Web Search}.
\newblock \emph{{IEEE Transactions on Knowledge and Data Engineering}},
  15\penalty0 (4):\penalty0 784--796, 2003.

\bibitem[Kleinberg(1999)]{authoritative_sources_in_a_hyperlinked_environment}
J.~Kleinberg.
\newblock {Authoritative Sources in a Hyperlinked Environment}.
\newblock \emph{{Journal of the ACM}}, 46\penalty0 (5):\penalty0 604--632,
  1999.

\bibitem[Kok and Brockett(2010)]{hitting_the_right_paraphrase_in_good_time}
S.~Kok and C.~Brockett.
\newblock {Hitting the Right Paraphrases in Good Time}.
\newblock In \emph{{Proceedings of the Annual Conference of the North American
  Chapter of the Association for Computational Linguistics}}, 2010.

\bibitem[Mei et~al.(2008)Mei, Zhou, and
  Church]{query_suggestion_using_hitting_time}
Q.~Mei, D.~Zhou, and K.~Church.
\newblock {Query Suggestion Using Hitting Time}.
\newblock In \emph{{Proceedings of the 17th ACM Conference on Information and
  Knowledge Management}}, 2008.

\bibitem[Norris(1997)]{markov_chains}
J.~Norris.
\newblock \emph{{Markov Chains}}.
\newblock Cambridge University Press, 1997.

\bibitem[Page et~al.(1999)Page, Brin, Motwani, and
  Winograd]{the_pagerank_citation_ranking_bringing_order_to_the_web}
L.~Page, S.~Brin, R.~Motwani, and T.~Winograd.
\newblock {The PageRank Citation Ranking: Bringing Order to the Web}.
\newblock Technical report, InfoLab, Stanford University, 1999.

\bibitem[Rajaraman and Ullman(2010)]{mining_of_massive_datasets}
A.~Rajaraman and J.~Ullman.
\newblock \emph{{Mining of Massive Datasets}}.
\newblock Cambridge University Press, 2010.

\bibitem[Sarkar and
  Moore(2007)]{a_tractable_approach_to_finding_closest_truncated_commute_time_neighbors_in_large_graphs}
P.~Sarkar and A.~Moore.
\newblock {A Tractable Approach to Finding Closest Truncated-commute-time
  Neighbors in Large Graphs}.
\newblock In \emph{{Proceedings of the 23rd Conference on Uncertainty in
  Artificial Intelligence}}, 2007.

\bibitem[Sarkar et~al.(2008)Sarkar, Moore, and
  Prakash]{fast_incremental_proximity_search_in_large_graphs}
P.~Sarkar, A.~Moore, and A.~Prakash.
\newblock {Fast Incremental Proximity Search in Large Graphs}.
\newblock In \emph{{Proceedings of the 25th International Conference on Machine
  Learning}}, 2008.

\bibitem[Yen et~al.(2005)Yen, Vanvyve, Wouters, Fouss, Verleysen, and
  Saerens]{clustering_using_a_random_walk_based_distance_measure}
L.~Yen, D.~Vanvyve, F.~Wouters, F.~Fouss, M.~Verleysen, and M.~Saerens.
\newblock {Clustering Using a Random Walk Based Distance Measure}.
\newblock In \emph{{Proceedings of the European Symposium on Artificial Neural
  Networks}}, 2005.

\bibitem[Zhu et~al.(2003)Zhu, Ghahramani, and
  Lafferty]{semi_supervised_learning_using_gaussian_fields_and_harmonic_functions}
X.~Zhu, Z.~Ghahramani, and J.~Lafferty.
\newblock {Semi-Supervised Learning Using Gaussian Fields and Harmonic
  Functions}.
\newblock In \emph{{Proceedings of the International Conference on Machine
  Learning}}, 2003.

\end{thebibliography}

\end{document}